# Phase resolved measurements of stimulated emission

Josef Kröll[1], Juraj Darmo[1*], Sukhdeep S Dhillon[3], Xavier Marcadet[4], Michel Calligaro[4], Carlo Sirtori[2], and Karl Unterrainer[1]

[1] *Photonics Institute, Vienna University of Technology, Gusshausstrasse 25-29, A-1040 Vienna, Austria*

[2] *Matériaux et Phénomènes Quantiques, Université Paris 7, 75251 Paris Cedex 05, France*

[3] *Ecole Normale Superieure, 75231 Paris Cedex 05, France*

[4] *Alcatel-Thales III-V Lab, Route Dépatementale 128, 91767 Palaiseau Cedex, France*

\* to whom correspondence should be addressed: e-mail: juraj.darmo@tuwien.ac.at

The development of the semiconductor quantum cascade laser (QCL) [1] has enabled bright coherent sources operating at frequencies between the optical (>100 THz) and electronic (<0.5 THz) ranges opening this frequency region for fundamental science investigations [2-5] as well as for applications [6]. However, detailed information about the internal processes in QCLs and their ultrafast dynamics are needed for their further development towards high output power, longer wavelengths and stable pulsed operation. We introduce a novel approach to study the gain and dynamics of a terahertz (THz) QCL [7] using phase resolved measurements of the stimulated emission. This is enabled by direct recording of the emitted electric field with <100 fs time resolution. For the case of the THz QCL we demonstrate the validity of the quantum mechanical description of the laser. We show for the first time the real bandwidth of the terahertz gain medium, the current dependence of this gain, and the losses associated with the wave propagation in the laser waveguide.

The dynamic behavior of lasers is of enormous importance for their application, detailed understanding and further improvements. Very attractive is e.g. the time window within the coherence time of the laser's optical transition. Another particular interest lies in the study of stimulated emission beyond the simple rate equations description in which the quantum mechanical processes are approximated by transition probabilities determined by Fermi´s Golden Rule. The perturbative approach here employed is quite simple and effective in simulating lasers. However, the insight to the underlying quantum mechanical processes is lost in this approach. A rigorous quantum mechanical description of stimulated emission shows that incoming photons generates a superposition between the ground state and the first excited state. This superposition creates the necessary oscillating dipole which emits electro-magnetic radiation. The difference between emission and absorption is only determined by the phase of the superposition. In the former case there is constructive interference between the incoming photon and the field emitted by the dipole resulting in amplification. In the later case, destructive interference between the incoming photon and the emitted electro-magnetic wave yields in an absorption. Since the realization of the first laser, researchers have investigated stimulated emission and the gain in an amplifying medium using coherent or incoherent radiation [8]. Several well know phenomena have been observed in these very basic experiments, i.e. the spectral gain curve, spectral narrowing, power broadening, and gain saturation [2-5]. However, a direct insight on the transient processes of stimulated emission has not been possible so far due to a necessity of phase resolved measurement of the electric field on the femtosecond time scale.

With the advent of ultrashort laser pulses, time-resolved spectroscopy with a time resolution better than 10 fs has become possible [9]. Phase-locked extreme UV pulse generation recently allowed probing of electronics dynamics in atoms even on the attosecond time scale [10]. At the same time, the femtosecond laser driven generation of few-cycle terahertz (THz) pulses has paved the way towards coherent detection of transient THz electric fields using electro-optic sampling [11]. This latest generation of phase-resolved femtosecond spectroscopy recently enabled a very close look at the formation of a plasmon-phonon mode [12] showing that the system requires a certain time for the transformation from a broadband to a narrowband response. Kaindl et al. used time-resolved THz spectroscopy to study the exciton formation dynamics [13].

The phase-resolved measurement of stimulated emission, however, has still been hindered by the rather fast electric field oscillations in conventional lasers (about 2.5 fs at 800 nm). The situation has changed dramatically with the realization of the THz quantum cascade laser (THz-QCL) [7]. This laser is based on a semiconductor heterostructure and the laser´s ground and first excited states are quantized energy levels of the heterostructure. The heterostructure related design freedom has allowed the fabrication of lasers emitting in a wide range of frequencies, from 4.4 THz down to 1.9 THz [7,14]. The required time resolution necessary to resolve the electric field oscillations of the THz-QCL emission was thus brought down to several hundreds of femtoseconds, which is readily accessible by modern femtosecond spectroscopy.

We use quasi single-cycle THz pulses generated from a semiconductor by excitation with femtosecond laser pulses [15] to study the interaction of the THz probe pulse with the population inverted energy level systems in the THz-QCL active region. The employment of electro-optic sampling of the THz electric field allows phase-resolved measurements of stimulated emission and terahertz pulse amplification in both amplitude and phase.

The interaction of a quasi single-cycle pulse with the optical transition in the QCL can be modeled via the interaction of an electromagnetic wave with a two-level quantum system. This approximation of a THz-QCL is adequate as the upper level of the two-level system (optical transition) is the lowest energy level of the injector region and the lower level of the two-level system is the highest energy level of the next injector region (Fig. 1c). We suppose that both energy levels interact with other levels in the heterostructure via tunneling and scattering at timescales much longer than the optical transition rate. Thus the two-level quantum system representing the laser's optical transition can be on a certain time scale considered as an isolated system. We use Maxwell-Bloch equations to numerically simulate the interaction of the THz pulse with this system. The results of these simulations are summarized in Fig 2. The electric field transient forces a coherence of the wavefunctions of the ground and excited states of the two-level quantum system and a temporal oscillating dipole is formed. The electric field emitted by this dipole is out of phase with the excitation field when the ground state population dominates (Fig. 2a). If a population inversion is present, the emitted electric field and the driving field are in phase (Fig.2b). The amplitude of the emitted field is proportional to the difference in population of the ground and excited states ($n_a$ - $n_b$) and the emission diminishes only for the case of $n_a = n_b$. The decay of the observed oscillations is governed by i/ a radiative decay of the population of the excited state, ii/ a non-radiative decay of the population of the ground and excited states $n_a$ and $n_b$ that relax to values given by thermal equilibrium, iii/ dephasing due to scattering, and iv/ dephasing within the inhomogeneously broadened ensemble of the two-level system.

The spectral content of the simulated response of the two-level system with an inverted population is show in Fig. 2c. The response consists of two components – the resonant main part corresponding to the eigenfrequency of the two-level system and the broadband part of the spectrum that corresponds to the instant response of two-level system to the driving electric field. In this later case, the population of the states is coherently driven by the electric field of the THz pulse and a partial Rabi oscillation is performed. To compare with the experiment, the propagation of the THz pulse through the gain medium is simulated using the finite-difference time-domain method applied to the Maxwell equations [16]. Figure 2d shows that, as expected, with increasing length of the pulse propagation in the gain medium, the oscillatory part of the pulse becomes dominant and eventually exceeds the intensity of the driving pulse injected into the QCL.

The uniqueness of our phase-resolved THz measurement method allows to observe the above discussed behavior of the laser gain medium directly in the time domain. The THz pulse transmitted through the unbiased THz-QCL laser is shown in Fig. 3a. The shape and time shift of the pulse is the result of the convolution of the probe pulse, the transfer function of the used THz coupling optics, and of the absorption in the unbiased THz-QCL. Namely, the transmitted pulse has the single cycle shape typical for a THz pulse generated from a biased semiconductor emitter [15], followed by damped oscillations with a period of about 350 fs. This temporal behavior points to a weak absorption in the THz-QCL waveguide at about 3 THz. When the laser is biased, the amplitude and the shape of the transmitted THz pulse change significantly. Fig. 3b shows the differential transmission of the THz pulse through the biased and unbiased THz-QCL. This time domain signal is dominated by a very well pronounced oscillation lasting several picoseconds after the probing pulse has passed. The oscillation represents the resonant response of the biased THz-QCL to the injected THz pulse due to stimulated emission.

The temporal shape of the oscillations is well explained by radiation of the dipole associated with our two-level model (see Figs. 3a and 2b). A detailed inspection of the relative phase between them and the electric field of the reference THz pulse shows that these fields are in phase (Fig. 3c), which means that we probe a two-level system (an optical transition in a THz-QCL) with an inverted population. This is in fact the very direct demonstration of the relevance of the quantum mechanical description of a laser. The amplitude of the oscillations decays exponentially with a time constant of about 7.5 ps and this decay time is independent from the driving current density. In principle, the oscillations decay is controlled either by the population lifetime on the optical transition, or by the dephasing intrinsic to the corresponding two-level system, or by a dephasing in the non-homogeneously broadened ensemble of the two-level systems. The later might be the most probable case for THz QCLs where non-uniformity of grown individual cascades (periods) and of redistribution of applied voltage are expected.

The amplitude and phase of the Fourier transformed THz-QCL modulation signal is shown in Figs 3d and 3e. The amplitude spectrum is dominated by a frequency component centered at 2.9 THz. This peak corresponds to the spectral gain of the THz-QCL which is also proven by the shape of phase spectrum. From the measurement we obtained an amplification coefficient of the electric field at ~2.9 THz of 8.5 which means the single pass gain is about 26 cm$^{-1}$. The bandwidth (FWHM) of the measured gain is about 130 GHz . The amplified frequency component correlates well with the emission of the THz-QCL, observed at 2.87 THz using an FTIR spectrometer.

The THz-QCL modulated signal contains also a broadband spectral feature at frequencies between 1.0 – 2.2 THz. The data analysis unambiguously shows that these spectral feature corresponds to a negative modulation (i.e. transmission decreases when the laser is on) that can be understood as increased losses. This modulation could be caused by a difference between the free carrier distribution in the biased and unbiased active region and related Drude-like response. However, the spectrum of the observed modulated signal cannot be fitted solely by Drude absorption, which would be proportional to the spectrum of the probing THz pulse (dashed line in Fig. 3d). This indicates that the confinement of the carriers in the heterostructures suppresses the Drude absorption in favor of an absorption by the several resonant intersubband transitions found within the THz-QCL injector region.

In Fig 4a we show the bias current dependence of the THz pulses transmitted through the THz QCL. At bias currents well below threshold the observed THz transient resembles the input THz pulse. The Fourier transformed spectral THz amplitude on the right hand side shows that there is reduced transmission for spectral components below 1.5 THz, and no amplification is observed at the THz QCL frequency. At this current density the cascades are not completely aligned and the lasing transition is not established yet. At threshold (picture in the middle, J = 113 A/cm2) the THz transient exhibits already a quite significant change. The initial THz pulse is now followed by several strong oscillations which, however decay quickly after the input pulse. The corresponding spectral amplitude shows now an interesting behaviour and demonstrates the potential of the phase resolved measurement. Below 2.5 THz the observed negative modulation relates to reduced transmission. Above 2.5 THz we have positive modulation and thus amplification. The spectral dependence of the amplified field is given by the gain curve of the QCL which seems to have become ideally aligned at that bias point. Only a close look on the phase allows separating unambiguously amplification from reduced transmission. The negative part at low frequencies is most probably caused by the population of the injector states which lie energetically quite close. Well above threshold (bottom panel in Fig.4) the THz transients are now completely dominated by the strong, long lasting oscillation. Form the area under the curve the amplification is obvious. The spectral dependence shows now strong amplification at the lasing transition of 2.9 THz. The negative modulation below 2.5 THz is also increased indicating that reabsorption by the injector state or even by the upper laser state to higher lying injector states is present.

The driving current dependence of the single pass gain at 2.9 THz (Fig. 5a) reveal details about the internal processes in the active region of the laser. Without an applied bias or with low bias, the THz-QCL exhibits no resonant absorption at the lasing frequency (~2.9 THz). This means that the active region of QCL switches at these frequencies from a transparency directly to gain. This somewhat untypical behavior of the laser gain medium is explained by the special nature of a quantum cascade laser. Without applied correct bias the energy levels in the QCL are not properly aligned [18], the given optical transition does not exist and no resonant absorption is observed. Our measurements show that a measurable gain appears when most of the 90 quantum cascades in the laser are properly aligned (at ~ 2.1 V). Some of the cascades, however, get aligned at quite low bias as is proven by the drop in the differential resistance of the laser (Fig. 5c). After the onset, measured gain rises proportionally with driving current density, which means that the inversion of electron population is increasing. At a bias current density of about 113 A/cm$^2$ the gain overcomes the laser's optical cavity losses and the laser action starts. The single pass gain value of ~ 19 cm$^{-1}$ observed at the threshold is consistent with expected waveguide and cavity mirrors losses. The waveguide loss of 12±1 cm$^{-1}$ was observed in the independent measurement performed on the unbiased THz-QCL waveguides of the same design but with different lengths. This value is in good agreement with that calculated theoretically and been of 7 – 11 cm$^{-1}$ [20-22], although the single till now published measurement of the waveguide loss gave value of 38±20 cm$^{-1}$ [18]. The remaining part of 7 cm$^{-1}$ of the laser losses corresponds to the losses of the cavity mirrors, which have combined reflectivity of about 20%. Such high loss is found due to the low reflectivity at the laser waveguide facet where the coupling hemispherical lens is placed.

In contrast to THz electric field resolving measurement the total CW output intensity as measured by a Golay cell detector shows only weak spontaneous emission below the threshold current density. At the threshold, the laser output power starts to increase with driving current density (Fig. 5b). The single pass gain of the THz pulse, however, does not increase linearly anymore in this current range, but it becomes almost "clamped" at the threshold value as expected according to the laser theory. We explain this fact by spatial gain clamping. The standing-wave pattern of the laser mode leads to a local gain clamping in the mode antinodes, while the gain is unclamped at the mode nodes. Therefore, the THz pulse passing through the laser cavity probes both areas with saturated and unsaturated gain. Since the mode's high intensity regions can occupy most of the active region, the slope of the single pass gain should be much smaller that below threshold. Actually, the slope of the gain is less than one eights of that below threshold suggesting an existence of very strong standing-wave pattern in the laser cavity. Regarding the quantitative values of the single pass gain, the maximum observed gain is estimated to be $28 \pm 2$ cm$^{-1}$.

At higher bias current densities ($> 180$ A/cm$^2$) the gain decreases (Fig. 5b), which is in the case of the THz-QCLs usually explained by a misalignment of the energy levels in individual quantum cascades when a critical electric field intensity is exceeded. Correspondingly, the observed single pass gain drops steeply and is associated with a steep increase of the laser differential resistance (Fig. 5c). However, the observed decrease of the laser output power (Fig. 5b) cannot be explained solely by misalignment of the whole structures since it starts at lower driving current density. There is also no indication for that from the current voltage characteristics since the kink in the output power characteristics of the laser does not correlate with the onset of the gain roll-off. Actually, the laser output power drops in spite of a sustained single pass gain. We explain that apparent discrepancy by a gradual increase of the laser's waveguide losses due to the temperature increase in the active region or due to high carrier injection at those driving conditions Lasing stops when the gain drops below a value that is now increased by $\sim 3$ cm$^{-1}$ compared to the threshold gain at 113 A/cm$^2$ which supports our explanation of the increased laser losses.

With the introduction of the phase-resolved measurement of gain media, a whole class of new experiments can be performed and important parameters can be directly accessed. The first issue is the unambiguous assignment of gain or loss by the sign of the phase. Since the complete dielectric response of the system is obtained, it is straightforward to separate out the effect of refractive index changes of the waveguide that in simple transmission measurements usually complicate the quantitative determination of the gain. This is especially important for high index waveguides and active media. The use of dual detection using an electric field resolving and an intensity sensitive detectors provides the basis for simultaneous measurements of amplification and self-sustained lasing. The individual access to these separate physical properties allows the determination of the waveguide or other optical losses, but nonlinear effects like saturation, gain clamping, and spatial gain modulation can be also studied. In the future this technique can be extended to pump and probe studies, where the response of the system to the initial probe pulse can be monitored and carrier and tunnelling times can be deduced. Another fundamental experiment would be the phase locking of the QCL by the THz pulses. This technique is not restricted to QCLs and it is applicable for any gain system and extendable into the infrared optical range. The knowledge of internal losses, of the true gain bandwidth as well as of the saturation intensities and dephasing times, is important for the improvement and further development of these lasers.

The authors acknowledge support by the Austrian Science Foundation (SFB ADLIS project) and by the EU project TeraNova (IST-511415).

**(Methods)**

THz time-domain spectroscopy system

We use the terahertz time-domain spectroscopy [19] to assess the optical properties of a THz quantum cascade laser. In our specific case, a broadband THz electromagnetic pulse is injected into the THz QCL waveguide through one of the laser's facets. The THz pulses are generated by a photoconductive switch [15] using 800nm, 80 fs short pulses from a Ti-sapphire mode-locked laser. The typical frequency spectrum of the THz pulse covers the range between $0.5 - 3.5$ THz (signal dynamic range >20 dB) with maximum at about 1.2 THz. The THz pulse coupled in the laser through one facet propagates along the laser cavity axis and is emitted at the second facet into free space. Electro-optic detection [11] is employed to sense the instantaneous electric field vector of the THz pulse. Electro-optic sampling is a unique technique that allows to resolve THz electric field oscillations in both magnitude and phase. It is a coherent detection technique so that only the radiation, which has constant phase shift with the sampling femtosecond pulse, is recorded. The time resolution of this detection technique is given by the length of the sampling near-infrared pulse (~80 fs) that is much shorter than the THz (< 10 THz) oscillation period. The amplitude and phase of spectral components contained in the measured time domain signal are obtained by standard Fourier transform technique and the typical frequency resolution of the set-up was 20 GHz. Complementary to the electro-optic detection of the THz electr4ic field we use also a Golay cell detector which records the total THz intensity. In order to suppress parasitic effects of scattered light and electrical disturbances in the set-up, the terahertz emitter and the laser were modulated at two different frequencies $\omega_1$ and $\omega_2$, and the measured signal was observed at difference frequency $\omega_1$-$\omega_2$.

The modulation THz signal is obtained as the difference between the electric field of the pulses transmitted through the laser with and without bias. The amplitude and phase changes of the THz pulse are measured at different driving current densities and are a direct evidence of the modified optical properties of the laser active region. This type of measurement principally differs from the standard measurements [2-5] performed on lasers till now. These differences are manifold and emphasize the *uniqueness* of the method:

1) high time resolution (<90 fs);

2) measurement over a broad frequency range compared to a narrow ranges of standard methods;

3) amplitude and phase sensitive measurement of the electric field of the amplified pulse in contrast to a time-averaged measurement of the intensity of a probing beam in standard methods.

THz Quantum Cascade Laser

The measurement is performed on a AlGaAs/GaAs terahertz quantum cascade laser with a bound-to-continuum optical transition [20,21]. The THz-QCL's active region consists of 90 cascades. The bandstructure of one period is (applied electric field) shown in Fig. 1c. The laser, processed into ridges of 170 μm width and 2 mm length, emits at 2.87 THz and operates at sink temperatures up to 90 K [21]. The typical dependence of the THz output power and the applied voltage on the driving current density is shown in Fig. 5b and 5c, respectively. The threshold current density is about 113 A/cm$^2$ and the lasing abruptly stops at the current density of 195 A/cm$^2$. All parameters refer to conditions of pulsed operation of the laser with the repetition rate of 17 kHz and the duty cycle of 15%. The laser was not capable of the CW operation mode due to excessive heating of the active region.

**References**

[1]   J. Faist, F. Capasso, D. Sivco, C. Sirtori, A. Hutchinson, and A. Cho, Quantum cascade laser, *Science* 264, 553-556 (1994).

[2]   W.R. Bennet Jr., Hole burning effect in a He-Ne optical maser, *Phys. Review* 126, 580-593 (1962).

[3]   W.W. Rigrod, Gain saturation and output power of optical masers, *J. Applied Physics* 34, 2602-2609 (1963).

[4]   R. Osgood, W. Eppers, andE. Nichols, An investigation of the high-power CO laser, *IEEE J. Quantum Electron.* QE-6, 145-154 (1970).

[5]   W. Crowe and W.F. Ahearn, Semiconductor laser amplifier, *IEEE J. Quantum Electron.* QE-2, 283-289 (1966).

[6]   D. Mittleman, Sensing with terahertz radiation (Springer, New York 2005).

[7]   R. Koehler, A. Tredicucci, F. Beltram, H.E. Beere, E.H. Linfield, A.G. Davies, D.A. Ritchie, R.C. Iotti, and F. Rossi, Terahertz semiconductor-heterostructure laser, *Nature* 417, 156-159 (2002).

[8]   M.O. Scully and W.E. Lamb, Quantum theory of an optical maser. I. General theory, *Phys. Review* 159, 208-226 (1967).

[9]   see e.g. J.-C. Diels and W. Rudolph, Ultrashort laser pulse phenomena: Fundametals, technique, and applications on a femtosecond time scale (Academic Press, San Diego 1996).

[10]  A. Baltuska *et al.*, "Attosecond control of electronic processes by intensive light fields", *Nature* 421, 611-615 (2003).

[11]  Q. Wu and X.-C. Zhang, Ultrafast electro-optic field sensor, *Appl. Phys. Lett.* **68**, 1604-1606 (1996).

[12]  R. Huber, F. Tauser, A. Brodschelm, M. Bichler, G. Abstreiter, and A. Leitenstorfer, How many-particle interactions develop after ultrafast excitation of an electron−hole plasma, Nature 414, 286-289 (2001).

[13]  R. Huber, R.A. Kaindl, B. A. Schmid, and D.S. Chemla, Broadband terahertz study of excitonic resonances in the high-density regime in GaAs/Al$_x$Ga$_{1-x}$As quantum wells, Phys. Rev. B 72, 161314 (2005).

[14]    S. Kumar, B.S. Williams, Q. Hu, and J.L. Reno, 1.9 THz quantum-cascade lasers with one-well injector, Appl. Phys. Lett. 88, 121123 (2006).

[15]    X.-C. Zhang, B.B. Hu, J.T. Darrow, and D.H. Auston, Generation of femtosecond eletromagnetic pulses from semiconductor surfaces, *Appl. Phys. Lett.* **56**, 1011-1013 (1990).

[16]    A. Taflove, Computational electrodynamics: The finite-difference time-domain method (Artech House, Boston 1995).

[17]    J. Faist, F. Capasso, C. Sirtori, D.L. Sivco, and A.Y. Cho, in Intersubband transitions in quantum wells: Physics and device applications II, eds. H.C. Liu and F. Capasso (Academic Press, London 2000).

[18]    M. Rochat, M. Beck, J. Faist, and U. Oesterle, Measurement of far-infrared waveguide loss using a multisection-single pass technique, *Appl. Phys. Lett.* 78, 1967-1969 (2001).

[19]    M.C. Nuss and J. Orenstein, Terahertz time-domain spectroscopy, in Millimeter and Submillimeter Wave Spectroscopy of Solids, ed. G. Gruener (Springer, Berlin 1998), Chap.2.

[20]    J. Alton, S.S. Dhillon, C. Sirtori, A. de Rossi, M. Calligaro, S. Barbieri, H.E. Beere, E.H. Linfield, and D.A. Ritschie, Burred waveguides in terahertz quantum cascade lasers based on two-dimensional plasmon modes, *Appl. Phys. Lett.* 86, 71109 (2005).

[21]    S. Barbieri, J. Alton, H. Beere, J. Fowler, E. Linfield, and D.A. Ritchie, 2.9 THz quantum cascade lasers operating up to 70 K in continuous wave, Appl. Phys. Lett. 85, 1674-1676 (2004).

[22]    S. Kohen, B. Williams, and Q. Hu, Electromagnetic modelling of terahertz quantum cascade laser waveguides and resonators, J. Appl. Phys. 97, 053106 (2005).

**(a)**

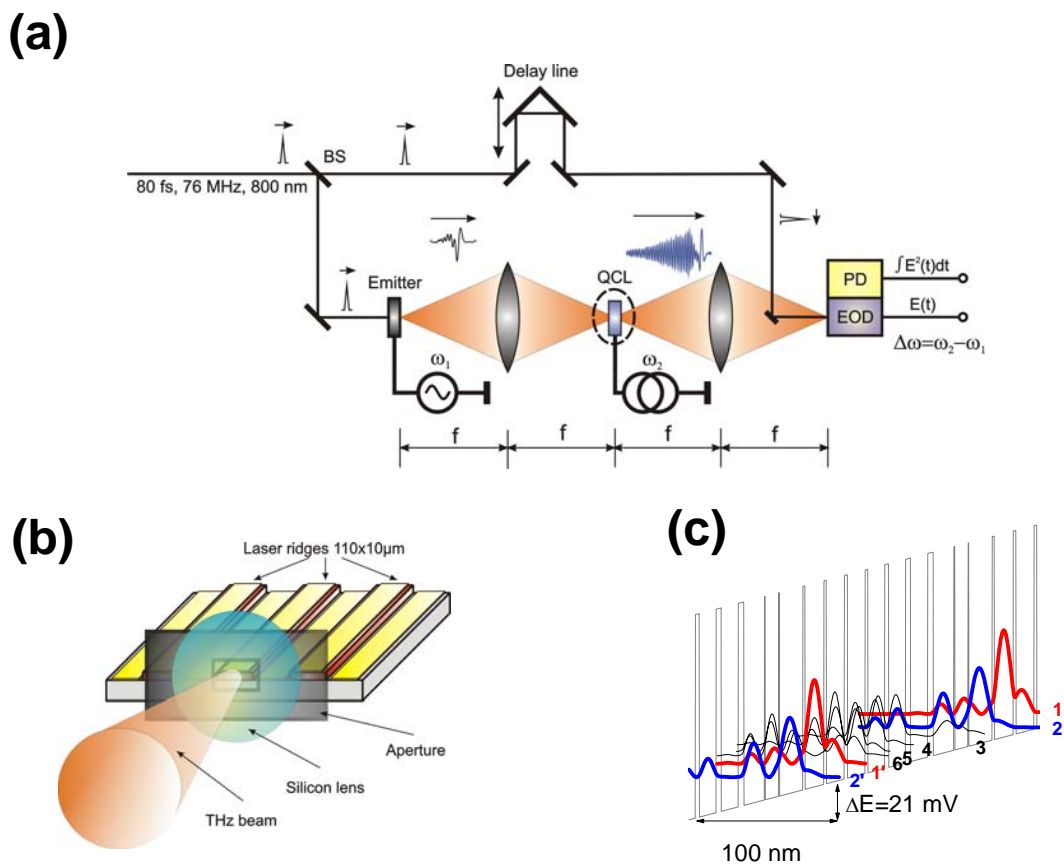

**(b)**

**(c)**

Fig.1 (a) Schematic of the measurement method: The source of femtosecond pulses is a Ti:Sapphire laser. The train of pulses is split by a beam splitter (BS) into two parts – one part is used to generate coherent THz pulses at the emitter and other for the sampling of the THz electric field at the electro-optic detector (EOD). Generated coherent THz pulses are transmitted through the quantum cascade laser and detected in the time domain using EOD. The average THz output power is measured using a Golay cell detector; (b) Sketch of the coupling of free-space THz waves into the THz-QCL waveguide: THz waves are focused by a primary lens onto a Si hemispherical secondary lens. The facet of the waveguide is placed at the beam waist of the hemispherical lens. The facet is shadow masked by a gold foil with a pinhole of dimensions 200x200 µm; (c) Band diagram of 1 period of the THz-QCL active zone. The active zone of the laser consists of 90 periods, each with a thickness 125 nm [20]. The dominant optical transition is between levels 1-2, while levels 3-6 form a transport channel for electrons between subsequent 1-2 groups. At the voltage drop of 21 mV per 100 nm of the active zone, the optical transition frequency is 2.9 THz. The laser processed into a ridge waveguide of 2 mm length and 170 μm width emits at 2.87 THz.

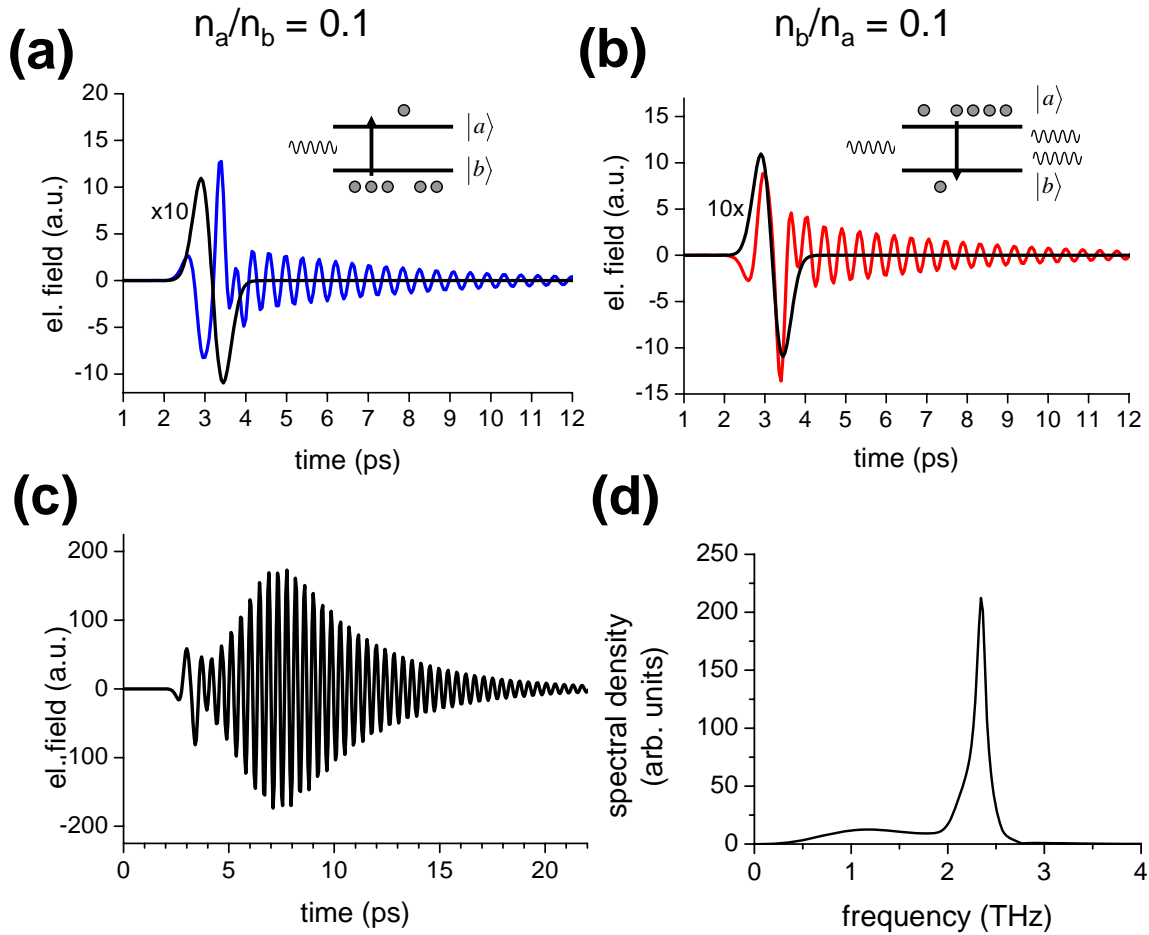

Fig.2   Numerical simulation of temporal response of the two-level system with (a) population in the thermal equilibrium $n_a/n_b$=0.1 (lossy system), and (b) the inversion of the population $n_b/n_a$=0.1 (system with a gain); (c) Electric field transient after propagation through the gain medium of the length of 2 mm and (d) corresponding spectrum. Simulation parameters: The homogeneously broadened two-level system, population $n_b/n_a$ = 0.1, transition frequency of 2.9 THz and linewidth of 0.12 THz. The driving THz pulse spectrum peaks at 0.75 THz.

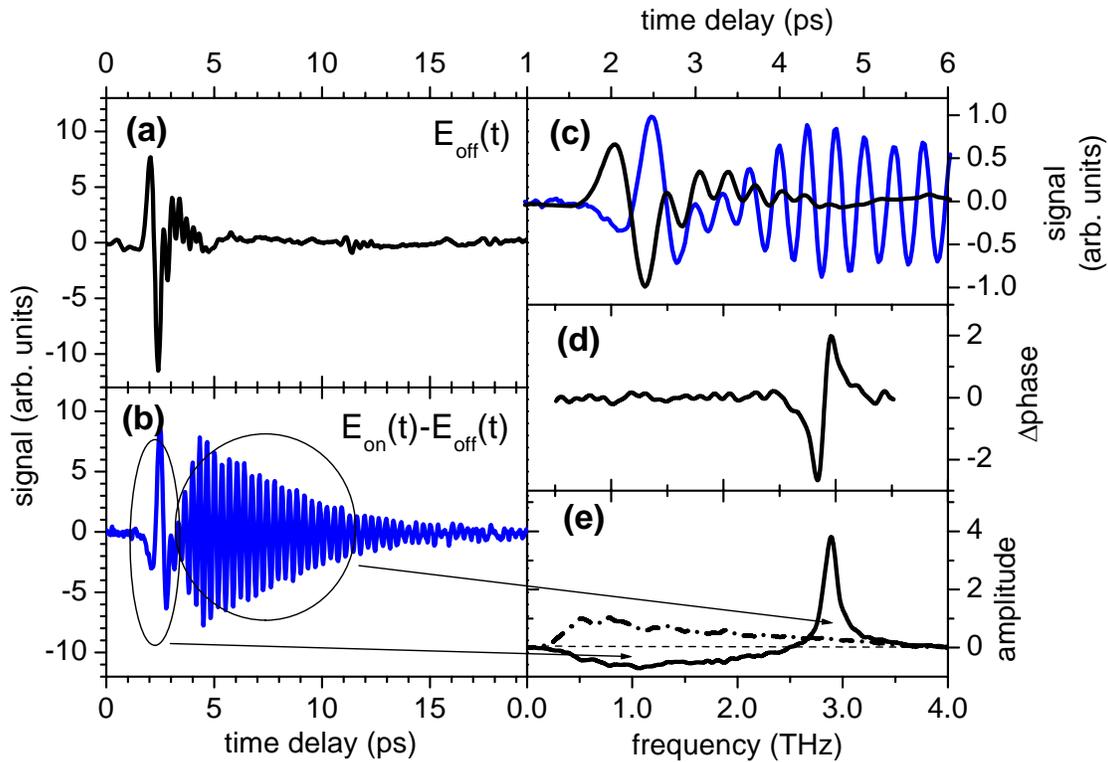

Fig.3   (a) THz pulse transmitted through the inactive THz-QCL; (b) Temporal shape of the emission from the THz quantum cascade laser initiated by the injected THz pulse; (c) Zoom of the normalized time domain signals of (a) and (b). The THz signals are phase shifted by π in the leading part and the oscillations have different period; (d) and (e): The amplitude and phase of the THz-QCL emission (solid lines). Since measured signal (b) is a mixture of changed transmission through the THz-QCL and laser emission, the spectrum contains negative (<2.5 THz) and positive (>2.5 THz) parts, respectively. The spectrum of the probing pulse is shown for comparison (dashed-dot line). During measurement the laser's heat sink temperature was 5 K and the laser's driving current density was 170 A/cm$^2$.

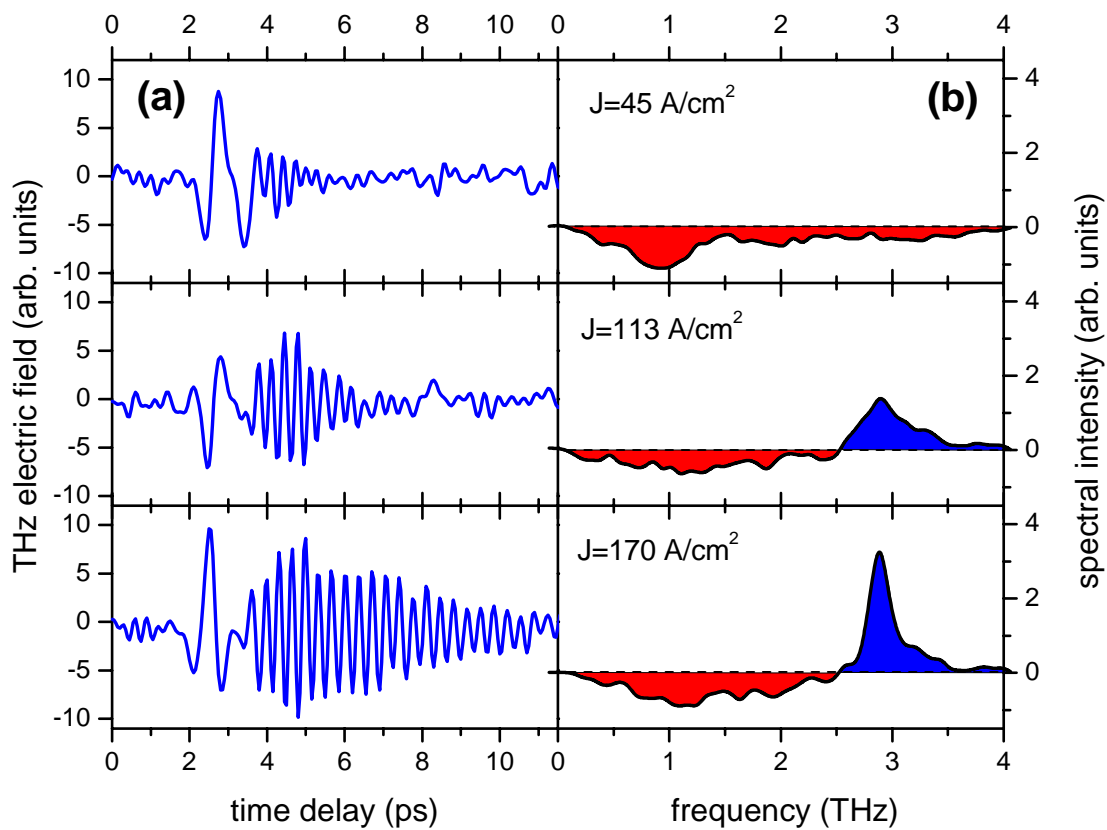

Fig.4. (a) Time domain and (b) corresponding spectral THz amplitude of the THz pulse transmitted through a THz-QCL at different bias current densities. The laser heat sink temperature was 5 K. In the upper picture the bias at the QCL is very low –no amplification is observed. The negative modulation at lower frequencies is due to reduced transmission. The picture in the middle shows the modulation right at threshold. The QCL cascades are aligned and a positive modulation signal shows amplification at the lasing transition. The increased negative modulation below 2.5 THz is explained by the population of the injector. In the lower picture the QCL is biased well above threshold; the transient signal is now dominated by strong oscillations which last quite long.

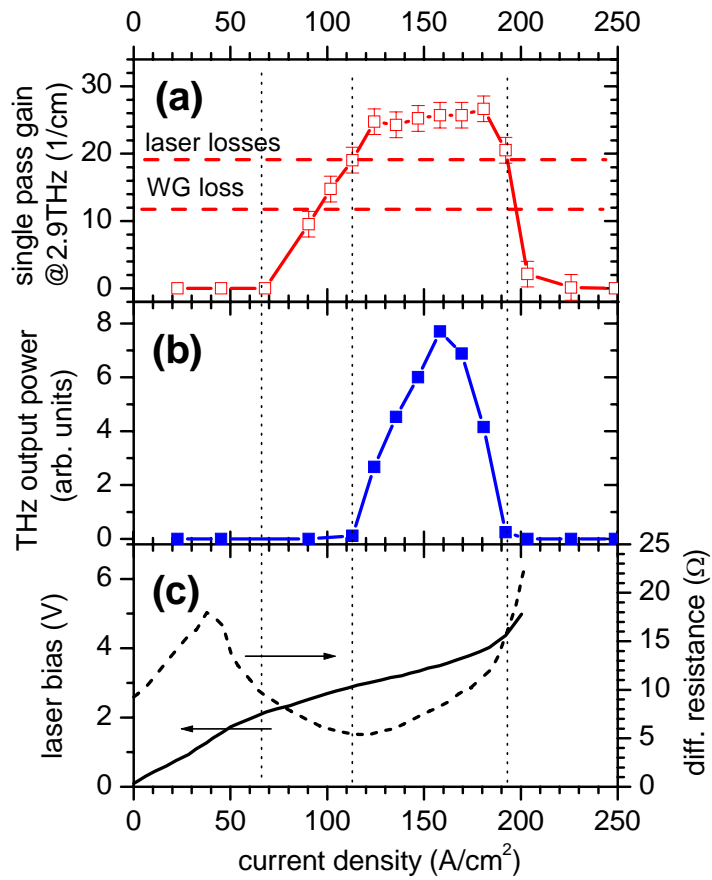

Fig.5   (a) Single pass gain as a function of the driving current density measured at 2.9 THz; (b) Laser average output power characteristics; (c) Bias (solid line) and differential resistance (dashed line) of the laser as a function of the driving current density. Laser was measured in the pulsed regime with a 17 kHz repetition rate and a 15% duty cycle. All measurements were done at a sink temperature of 5 K.